\documentclass{optica-article}
\journal{opticajournal} 
\articletype{Research Article}

\usepackage{lineno}
\usepackage{braket}

\usepackage{bm}

\begin{document}

\title{Intensity correlation holography for remote phase sensing and 3D imaging}

\author{Guillaume Thekkadath,\authormark{1,*} Duncan England,\authormark{1} and Benjamin Sussman\authormark{1,2}}

\address{\authormark{1}National Research Council of Canada, 100 Sussex Drive, Ottawa, K1N 5A2, Canada\\
\authormark{2}Department of Physics, University of Ottawa, Ottawa, K1N 6N5, Canada}

\email{\authormark{*}guillaume.thekkadath@nrc.ca}

\begin{abstract*} 
Holography is an established technique for measuring the wavefront of optical signals through interferometric combination with a reference wave.
Conventionally the integration time of a hologram is limited by the interferometer coherence time, thus making it challenging to prepare holograms of remote objects, especially using weak illumination.
Here, we circumvent this limitation by using intensity correlation interferometry.
Although the exposure time of individual holograms must be shorter than the interferometer coherence time, we show that any number of randomly phase-shifted holograms can be combined into a single intensity-correlation hologram.
In a proof-of-principle experiment, we use this technique to perform phase imaging and 3D reconstruction of an object at a $\sim$3~m distance using weak illumination and without active phase stabilization.
\end{abstract*}

\section{Introduction}
Determining the phase distribution of a wave from a set of intensity measurements is a fundamental problem in physics. 
In optics, it is a key component in various imaging techniques, including quantitative phase imaging, holography, and ptychography~\cite{schnars2015digital,pfeiffer2018x,park2018quantitative,nguyen2022quantitative}. 
It also finds applications in astronomy and free-space communications, where phase aberrations in a wavefront can be corrected using adaptive optics~\cite{tyson2022principles}.

Optical wavefronts are generally characterized using some sort of interferometric technique.
Self-referencing techniques, such as wavefront sensors~\cite{platt2001history,wu2019wish}, shear plates~\cite{bates1947wavefront}, and "direct" methods~\cite{bamber2012measurement}, either interfere the signal wave with itself or perform local calibrated phase transformations.
These have less stringent stability requirements, but can exhibit limitations such as reduced spatial resolution or ambiguities around amplitude null points~\cite{fried1998branch}.
Alternatively, holographic techniques involve interfering the signal with an external reference wave. 
The use of a reference introduces stability requirements, but alleviates null point issues and can enhance the measurement signal-to-noise ratio (SNR) through heterodyne gain~\cite{thornton2019deep,wolley2023near}. 

Conventionally, a hologram is obtained in the following manner. 
A single light source is split into two paths within an interferometer. 
One path prepares the signal wave, e.g. by illuminating an object, while the other path serves as the reference. 
The two paths are then recombined, and the hologram is recorded by measuring the amplitude (first-order) interference pattern, also known as the interferogram. 
The phase distribution of the signal wave is typically retrieved from multiple interferograms acquired with varying phase shifts by tuning the interferometer path length~\cite{yamaguchi1993phase} or using an off-axis configuration~\cite{schnars1994direct}. 
However, these approaches require that the interferometer is stable for the entirety of the measurement.
Noise such as vibrations limits the hologram integration time, which in turns limits the precision of the retrieved phase.
Increasing the illuminating power can enhance the precision, but this approach may not always be suitable~\cite{gross2007digital,marim2011off} .
For example, in remote holography, diffuse reflection and poor interferometric stability necessitate the use of high-energy pulsed lasers~\cite{goodman1969experiments}.
Moreover, one may be constrained to using low powers when imaging samples susceptible to photodamage~\cite{charriere2007influence,zhang2020holo,mirecki2022low}, particularly with shorter wavelengths.

While the above techniques are based on amplitude interference, it is also possible to perform holography using intensity (second-order) interference by measuring intensity correlations between the signal and reference fields~\cite{beard1969imaging,naik2011photon,wang2013experimental,kumar2014recovery,takeda2014spatial,huang2020measuring}.
Such correlations can reveal coherence properties in stochastically-varying fields~\cite{brown1956test}.
In particular, an intensity-correlation hologram can be made insensitive to temporal fluctuations in the phase offset between the signal and reference, provided that the correlation time window is sufficiently short.
Thus, the hologram can be integrated for arbitrarily long without having to stabilize the interferometer.
This idea was demonstrated in Refs.~\cite{thekkadath2023intensity,szuniewicz2023noise} using an intensified CMOS camera to perform holography at the single-photon level.
In this work, we show that this idea can also be achieved in a conventional holography setup.
Intensity correlations are extracted from interferograms captured by a standard frame-based camera.
Although the exposure time of each individual frame must be shorter than the interferometer coherence time, an arbitrarily large number of randomly phase-shifted interferograms can be combined into a single intensity-correlation hologram, thereby enabling integration times beyond the coherence time.
The signal wavefront can be then be retrieved using a principal component analysis on the intensity correlation hologram.
Building on Refs.~\cite{thekkadath2023intensity,szuniewicz2023noise}, our work presents two main novel results.
Firstly, we show that the technique works even when the interferograms are dominated by camera readout noise, thus making it applicable to a broad range of cameras.
Secondly, we use digital holographic interferometry to perform 3D imaging of a scattering object located at a distance of $\sim$3~m. 
By eliminating the need to actively phase-lock the interferometer, our technique facilitates holography applications restricted to a weak photon flux or a short coherence time, such as phase-sensitive microscopy of fragile samples or sensing of remote objects.

\section{Concept}
The signal and reference waves are combined and a camera captures a time series of interferograms $\{f_i(r)\}$, where $i$ is a frame index. 
Each interferogram is described by
\begin{equation}
    f_i(r) = A_i(r)\cos[\phi(r) + \theta_i]  + B_i(r)
    \label{eqn:interferogram}
\end{equation}
where $r=(x,y)$ is the pixel coordinate, $A(r)$ is the modulation depth of the interference pattern which depends on the signal and reference intensities, while $B(r)$ is a background term which includes readout and dark current noise from the camera.
The quantity $\phi(r)$ is a temporally-fixed spatial phase and is the central quantity of interest.
It describes the relative local phase variation between the signal and reference wavefronts, $\phi(r) = \phi_{\mathrm{sig}}(r) - \phi_{\mathrm{ref}}(r)$.
We assume that the reference has a spatially-uniform wavefront ($\phi_{\mathrm{ref}}(r)=0$) such that $\phi(r)$ is simply the wavefront of the signal wave.
In contrast, $\theta_i$ is a global phase offset between the signal and reference waves, which fluctuates between $[0, 2\pi)$ on a timescale of $\tau_c$, the interferometer coherence time.
These fluctuations are typically caused by changes in the interferometer path lengths due to vibrations and other environmental noise.
Importantly, the camera exposure time $t_e$ must be shorter than $\tau_c$ to ensure that $\theta_i$ is fixed in each interferogram.

In the low-light limit, such as when an object is far away, we need to combine many interferograms in order to achieve a sufficient SNR. 
However, since $\theta_i$ is randomly varying, we cannot simply add these intereferograms together as $\sum_i f_i(r)$ would ``wash-away" the interference term. 
While there are a number of algorithmic techniques~\cite{okada1991simultaneous,wang2004advanced,hao2009random,vargas2011phase,xu2011phase} to retrieve $\phi(r)$ from a series of randomly-shifted interferograms, these approaches are not the most efficient solution to the problem.
Instead, here we propose measuring intensity correlations between pixels in the interferograms rather than only the intensities of individual pixels. 
As we show mathematically below, this allows us to combine the interferograms in a way that is intensive to drifts in global phase.
Intensity correlations between pairs of pixels are given by:
\begin{equation}
\begin{split}
\braket{G(r,r')} &= \sum_i f_i(r)f_i(r') \\
&= \frac{1}{2}\braket{A(r)}\braket{A(r')}\cos{[\phi(r) - \phi(r')]} + \braket{B(r)}\braket{B(r')}.
\end{split}
\label{eqn:intensity_correlation}
\end{equation}
The bracket $\braket{}$ denotes an average over inteferograms (i.e. frames). 
We used the trigonometric identity $\cos{[a]}\cos{[b]} = (\cos{[a-b]}+\cos{[a+b]})/2$, and the fact that $\braket{\cos{[\phi(r) + \theta_i]}}=0$ since $\theta_i$ is randomly drifting between $[0, 2\pi)$ between interferograms. 
In quantum optics, Eq.~\eqref{eqn:intensity_correlation} is known as Glauber's second-order cross-correlation function.
It describes intensity correlations between the signal and reference waves.
Due to the factor of $1/2$ in the modulation term, the visibility of intensity interference is limited to 50\%~\cite{ou1988quantum}.
The benefit is that $\braket{G(r,r')}$ does not depend on the phase offsets $\theta_i$, but does depend on the quantity of interest, $\phi(r)$.
This key fact enables one to measure $\braket{G(r,r')}$ with arbitrarily long integration times without having to stabilize the interferometer.
The intensity-correlation hologram $\braket{G(r,r')}$ can be measured with an event-based camera by making a histogram of pairs of pixels that fired within $\tau_c$~\cite{thekkadath2023intensity}, or with a frame-based camera by combining frames captured with an exposure time $t_e < \tau_c$, as we show experimentally below.

Intensity-correlation holography [Eq.~\eqref{eqn:intensity_correlation}] presents a different phase-retrieval problem than Eq.~\eqref{eqn:interferogram}:
(i) there is a single input to the problem, which is the intensity correlation function averaged over all interferograms,
(ii) the problem is self-referencing, since the phase offsets are now given by $\phi(r')$, the quantity of interest, rather than an external $\theta_i$.
Using a trigonometric identity, we can re-write Eq.~\eqref{eqn:intensity_correlation} as
\begin{equation}
\begin{split}
\braket{G(r,r')} &= \frac{1}{2}\braket{A(r)}\braket{A(r')} \left( \cos{[\phi(r)]}\cos{[\phi(r')]} + \sin{[\phi(r)]}\sin{[\phi(r')]}\right) \\ 
&\quad + \braket{B(r)}\braket{B(r')} \\
&= f_{\mathrm{bg}}(r)f_{\mathrm{bg}}(r') + f_{\mathrm{cos}}(r)f_{\mathrm{cos}}(r') + f_{\mathrm{sin}}(r)f_{\mathrm{sin}}(r'),
\end{split}
\label{eqn:intensity_corr_ortho}
\end{equation}
where
\begin{equation}
\begin{split}
f_{\mathrm{bg}}(r) &= \braket{B(r)} \\
f_{\mathrm{cos}}(r) &= \braket{A(r)}\cos{[\phi(r)]}/\sqrt{2}  \\
f_{\mathrm{sin}}(r) &= \braket{A(r)}\sin{[\phi(r)]}/\sqrt{2}.
\end{split}
\label{eqn:components}
\end{equation}
If the phase distribution $\phi(r)$ takes on values evenly distributed between $[0, n\pi)$, for any integer $n$, then 
\begin{equation}
\sum_{r} \cos{[\phi(r)]}\sin{[\phi(r)]}  \approx 0,
\label{eqn:approx}
\end{equation}
and the components in Eq.~\eqref{eqn:components} are approximately orthogonal. 
This approximation holds when the inteferograms contain both destructive and constructive interference in different parts of the image, e.g. a fringe pattern.
In this case, the components in Eq.~\eqref{eqn:components} can be determined via an eigendecomposition of $\braket{G(r,r')}$.
They can also be determined by performing a principal component analysis (PCA) of a data matrix containing all the interferograms.
If we define $\bm{F} = [\bm{f}_1,...,\bm{f}_K]$, where $\bm{f}_k$ is a column vector corresponding to the flattened $k$th interferogram [Eq.~\eqref{eqn:interferogram}], then $\bm{G}=\bm{F}\bm{F}^T$ is equivalent to $\braket{G(r,r')}$ (where $^T$ denotes the transpose).
The principal components of $\bm{F}$ (and thus the eigenvectors of $\bm{G}$) can be found directly by performing a singular value decomposition (SVD) of $\bm{F}$.
The phase can then be retrieved using 
\begin{equation}
\phi(r) = \arctan{\left(\frac{f_{\mathrm{sin}}(r)}{f_{\mathrm{cos}}(r)}\right)},
\end{equation}
as outlined in Table~\ref{table:algo}. 

\begin{table}
\centering
\begin{tabular}{|l|} 
 \hline
 1. Define $\bm{F} = [\bm{f}_1,...,\bm{f}_K]$ where $\bm{f}_k$ is a column vector corresponding \\
 ~~~~to the flattened $k^{\mathrm{th}}$ interferogram. \\ 
 2. Perform a singular value decomposition of $\bm{F}$ such that $\bm{F}=\bm{U}\bm{\Sigma} \bm{V^T}$.\\
 3. Compute the phase via $\bm{\phi} = \arctan{(\bm{U}_3/\bm{U}_2)}$ where $\bm{U}_3$ ($\bm{U}_2$) is \\
 ~~~~the third (second) column vector of $\bm{U}$.\\
 4. Unflatten the retrieved phase $\bm{\phi}$. \\ 
 \hline
\end{tabular}
\caption{Phase-retrieval algorithm.}
\label{table:algo}
\end{table}

Our procedure is related to the phase-retrieval algorithm presented in Ref.~\cite{vargas2011phase}.
That work considers an eigendecomposition of the intensity correlation matrix averaged over pixels (i.e. correlations \textit{between frames}), whereas our method considers an eigendecomposition of the intensity correlation matrix averaged over frames (i.e. correlations \textit{between pixels}). 
The more computationally efficient method depends on how $K$ compares in size to $ N_x \times N_y$, i.e. performing eigendecomposition of $\bm{F}\bm{F}^T$ (size $ N_xN_y \times N_xN_y$) or $\bm{F}^T\bm{F}$ (size $K \times K$).
Since only the first few eigenvectors with the largest eigenvalues are needed, these can be efficiently found using partial SVD methods such as "locally optimal block preconditioned conjugate gradient" (LOBPCG) directly on the matrix $\bm{F}$, without having to construct either correlation matrix.
We used the Python implementation \texttt{scipy.sparse.linalg.svds}.
In contrast to Ref.~\cite{vargas2011phase}, we do not perform any background subtraction before performing the PCA.
Instead, we use the PCA itself to filter the background term $f_{\mathrm{bg}}$ as we found this approach improves the accuracy of the retrieved phase when the frames are dominated by read noise.
In practice, $f_{\mathrm{bg}}$ has the largest singular value due to noise and non-perfect interference visibility.
Then, $f_{\mathrm{cos}}$ and $f_{\mathrm{sin}}$ have the second and third largest singular values.
The relative ordering between these two components is a priori not known and so $\phi(r)$ is retrieved up to a global offset.

\section{Results}
\subsection{Reflective object: phase imaging of a spatial light modulator}
\label{sec:slm}

\begin{figure}
\centering
\includegraphics[width=1\columnwidth]{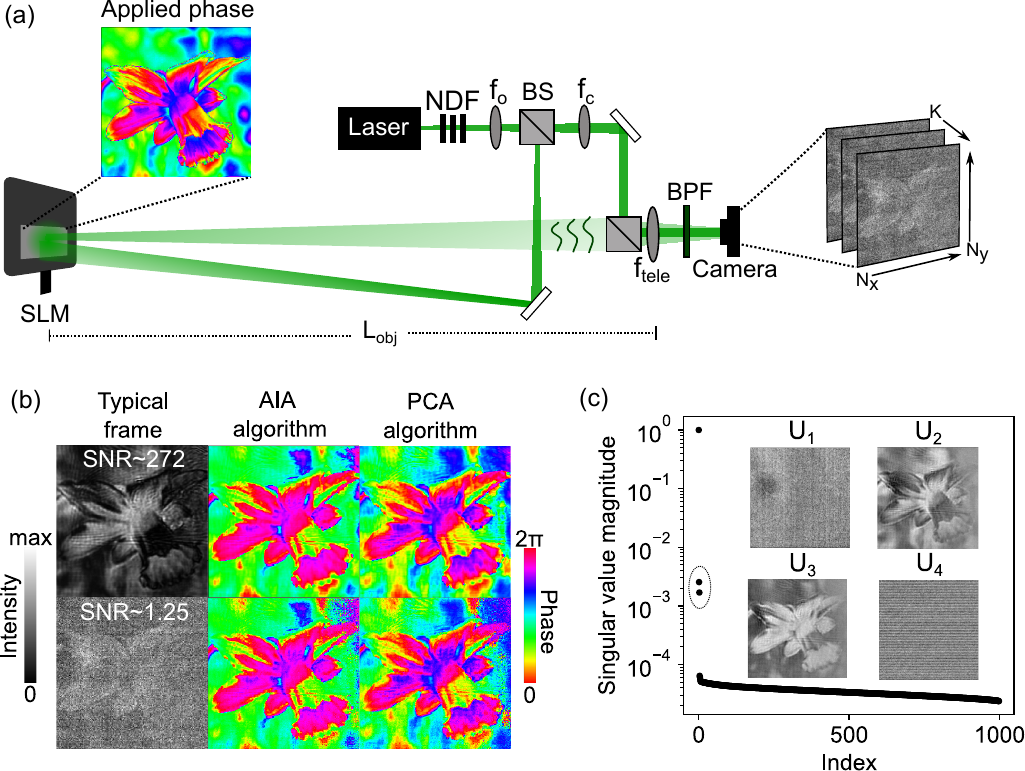}
\caption{\textbf{Holography of a reflective object}.
(a) Experimental setup. NDF: neutral density filter, BPF: bandpass filter, BS: beam splitter, SLM: spatial light modulator. 
(b) Retrieved phase using the principal component analysis (PCA) algorithm described in the main text and the advanced iterative algorithm (AIA) of Ref.~\cite{wang2004advanced}.
We use $K=1000$ randomly phase-shifted interferograms such as the ones shown in the column labelled ``typical frame".
Top (bottom) row shows results when interferograms have a high (low) SNR.
(c) Distribution of the singular value magnitudes from the PCA in the low SNR case.
Insets show the first four principal components, where $U_1 = f_{\mathrm{bg}}$, $U_2 = f_{\mathrm{cos}}$, $U_3 = f_{\mathrm{sin}}$, and $U_4$ is the largest noise component. 
}
\label{fig:reflective_results}
\end{figure}

We test the phase retrieval algorithm using the experimental setup shown in Fig.~\ref{fig:reflective_results}(a).
The light source is a continuous-wave laser with a 532 nm central wavelength and $>$ 300 m coherence length (CrystaLaser CL532-100-S).
We use a spatial light modulator (SLM, Meadowlarks E-Series 1920×1200) as a reflective phase object.
The SLM can be programmed to apply an arbitrary phase mask $\phi(r)$ within $[0, 2\pi)$ to the signal beam.
An objective lens ($f_o = 3$~mm) is used to expand the illuminating beam, while a second collimating lens ($f_c = 300$~mm) is placed in the reference path to ensure that the reference has a flat wavefront.
The interferometer is purposely made large ($L_{\mathrm{obj}} = 3.14$~m) to mimic a remote phase sensing scenario.
Its coherence time is approximately $\tau_c \sim 10$~ms.
A telescope lens ($f_{\mathrm{tele}}=1000$~mm, diameter 50.8~mm) images the SLM plane onto the camera (UI-3240CP-NIR-GL).
We record interferograms of size $N_x = N_y = 230$ pixels (pitch $p=5.3~\mu$m), corresponding to a size of $2.4$ mm $\times$ $2.4$ mm in the SLM plane.
The camera has a maximum frame rate of $R=60$ frames-per-second, quantum efficiency of $\sim70$\%, and readout noise of 88(1) electrons/pixel (root-mean-square).
Given that we use exposure times of $t_e \leq 1$~ms, the total hologram acquisition time $T$ is mainly determined by the deadtime between frames, i.e. $T \sim K/R$.
A bandpass filter (Semrock LL01-532) is used to reduce the contribution of room lights to the background levels.

We begin by testing the algorithm using $t_e = 0.2$~ms and a relatively high power of 3.4~$\mu$W for the combined signal and reference beams.
As shown in the top row of Fig.\ref{fig:reflective_results}(b), the resulting interferogram has a SNR of $\sim 272$, which is determined by the ratio of signal to readout noise electrons in each pixel. 
We capture $K=1000$ interferograms, each with a randomly-varying phase offset due to vibrations and other instabilities in the interferometer.
Combining these interferograms in the PCA algorithm, we can retrieve the SLM phase $\phi(r)$, as expected.
Remarkably, when we lower the power to $15.6~$nW such that each interferogram contains significant readout noise (SNR $\sim 1.25$), we still retrieve $\phi(r)$ with similar accuracy [bottom row, Fig.\ref{fig:reflective_results}(b)].
This suggests that the quality of the retrieved phase is limited by systematic errors such as the SLM calibration rather than the SNR of each interferogram.  
The PCA effectively isolates signal intensity correlations from the noise, as shown by the singular value distribution in Fig.~\ref{fig:reflective_results}(c).
We also compare our approach to the advanced iterative algorithm (AIA) of Ref.~\cite{wang2004advanced}.
While the AIA algorithm is able to retrieve the phase from the $K$ interferograms, the PCA algorithm shows better performance.
It is also much faster, taking only 1~s on a regular laptop whereas the AIA takes about 100~s.

\begin{figure}
\centering
\includegraphics[width=1\columnwidth]{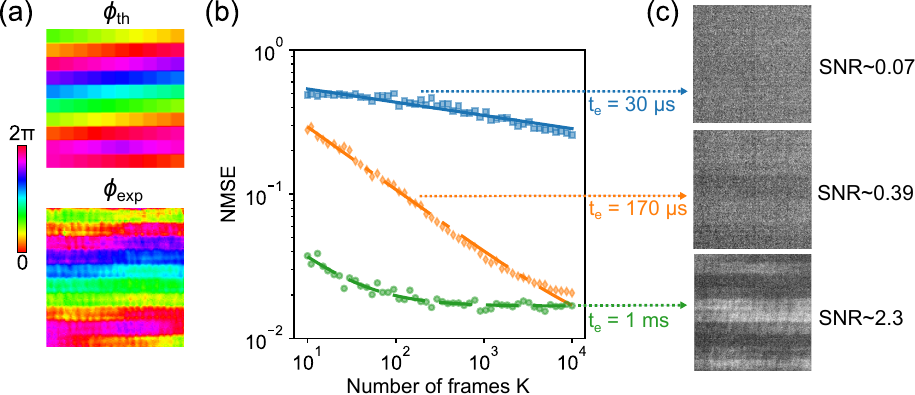}
\caption{\textbf{Phase error}.
(a) Applied $\phi_{\mathrm{th}}$ and an example of the retrieved $\phi_{\mathrm{exp}}$ (with $K=10,000$ and $t_e = 1~$ms) phase patterns used for the error analysis. 
(b) Log-log plot of the normalized mean square error (NMSE) of the retrieved phase using $K$ frames.
Each curve corresponds to a different exposure time $t_e$.
(c) Typical frame for each exposure time. 
The SNR of the interferograms is obtained by dividing the number of signal photoelectrons by the read-noise photoelectrons.
}
\label{fig:phase_error}
\end{figure}

Next, we fix the combined signal and reference power to 5.7 nW, resulting in approximately $200$ signal electrons/pixel/ms. 
We then vary $t_e$ and quantify the PCA performance by computing the normalized mean square error (NMSE) between the experimentally retrieved phase pattern $\phi_{\mathrm{exp}}(r)$ and the applied phase pattern $\phi_{\mathrm{th}}(r)$:
\begin{equation}
\mathrm{NMSE} = \frac{1}{\mathcal{N}} \sum_r |\phi_{\mathrm{exp}}(r) - \phi_{\mathrm{th}}(r)|^2 
\end{equation}
where $\mathcal{N} = \sum_r |\phi_{\mathrm{random}}(r) - \phi_{\mathrm{th}}(r)|^2$ is the mean squared error for a random phase pattern $\phi_{\mathrm{random}}(r)$.
The NMSE measures the accuracy of the retrieved phase relative to a random guess ($\mathrm{NMSE}=1$).
The results are shown in Fig.~\ref{fig:phase_error}.
We fit the NMSE to $\sqrt{(\alpha/K^x)^2 + \beta^2}$, which adds two noise terms in quadrature.
The first term, $\alpha/K^x$, models noise which reduces with the number of frames $K$, such as shot-noise ($x=1/2$).
The second term, $\beta$, models noise that is constant with $K$, such as readout noise, dark noise, and other systematic errors.
For $t_e = 30~\mu$s [blue curve], the interferograms are dominated by readout noise.
We find $x=0.09(2)$ (uncertainty is one standard deviation), and thus the phase-retrieval accuracy only benefits slightly from increasing $K$.
For $t_e = 170~\mu$s [orange curve], the frames still contain significant readout noise, and we find $x=0.43(1)$.
This performance nearly scales with shot-noise, which suggests that the PCA algorithm is effectively eliminating the camera noise from the interferograms.
Finally, for $t_e = 1$~ms [green curve], the signal exceeds the camera noise level and we obtain $x=0.49(4)$, in close agreement with shot-noise.

\subsection{Scattering object: 3D imaging of a dice}
\begin{figure}
\centering
\includegraphics[width=0.85\columnwidth]{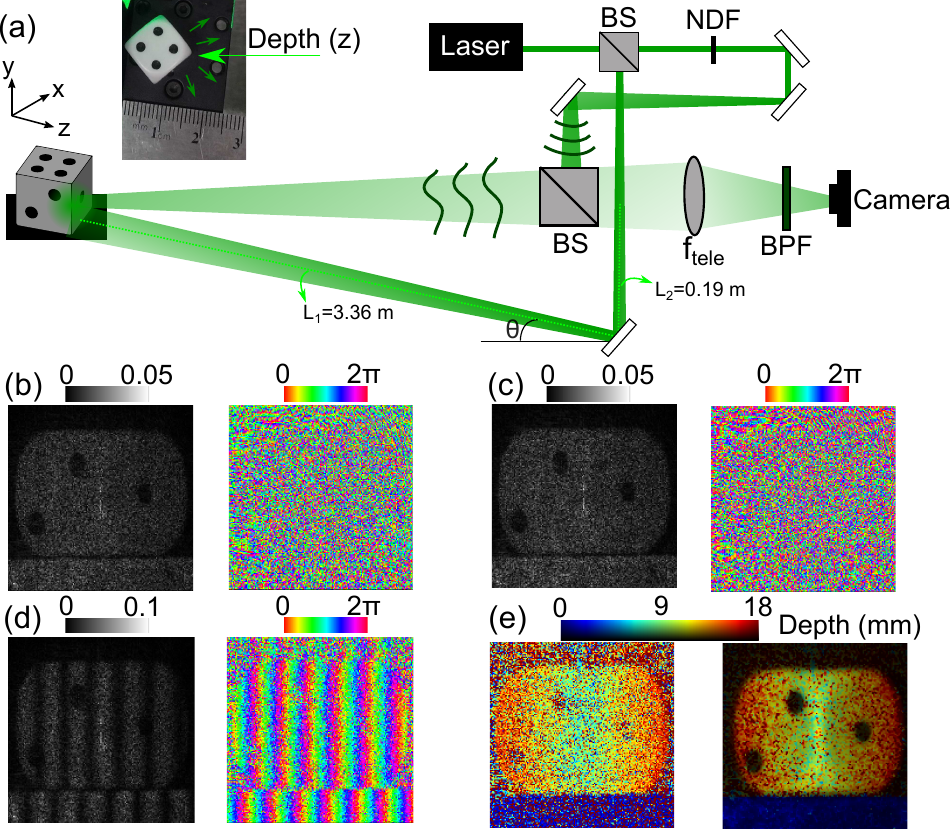}
\caption{\textbf{Holography of a scattering object}.
(a) Experimental setup. Inset is a photo of the object seen from above (i.e. looking down the y axis).
(b) and (c) Reconstructed intensity (left) and phase (right) distributions with illumination angle $\theta_1$ and $\theta_2$.
(d) Interference pattern $I(x,y)(1+\cos{(\Delta \phi)})$ (left) and phase difference map $\Delta \phi$ (right).
(e) Left shows the recovered depth map without any filtering.
Right shows depth map with the colormap modulated by the intensity $I(x,y)$ and applying a median filter to reduce speckle noise.
}
\label{fig:scattering_results}
\end{figure}

We also investigate the phase-retrieval algorithm on interferograms obtained using a scattering object.
As shown in Fig.~\ref{fig:scattering_results}(a), a 15 mm $\times$ 15 mm $\times$ 15 mm dice is mounted on a platform that is placed $L_{\mathrm{obj}}=3.97$~m away from a telescope lens with $f_{\mathrm{tele}}=500$~mm and a diameter of 50.8~mm (NA $\sim$ 0.05).
Using 125 mW to illuminate the object, we collect roughly 1 mW of scattered light.
A neutral density filter is used in the reference path to attenuate the reference to a similar power level.
All results below are obtained using $K=200$ frames and a $t_e =3$~ms exposure time.
Due to the maximum frame rate of the camera (60 frames per second), the total hologram acquisition time is $T\sim 3.3$~s. 

We first reconstruct $E_1 = \sqrt{I(x,y)}e^{i\phi_1(x,y)}$ when the object is illuminated at an angle $\theta$.
The phase $\phi_1(x,y)$ is determined using the algorithm outlined in Table~\ref{table:algo}, whereas $I(x,y)$ is measured by blocking the reference beam.
Both quantities display a speckle pattern [Fig.~\ref{fig:scattering_results}(b)].
While the speckles may look like noise, they are in fact real features in $E_1$ resulting from light scattering from the rough surface of the dice. 
We demonstrate the coherence of the speckle pattern using digital holographic interferometry (DHI)~\cite{schnars2015digital}.
We tilt the bottom mirror by a small amount $\Delta \theta$ along the $x$ direction, and again reconstruct the field $E_2 = \sqrt{I(x,y)}e^{i\phi_2(x,y)}$ [Fig.~\ref{fig:scattering_results}(c)].
Taking the difference between the reconstructed fields, $|E_1 - E_2|^2$, reveals an interference pattern [Fig.~\ref{fig:scattering_results}(d)].
This operation cancels out the phase shifts in the scattered wavefronts resulting from the rough dice surface.
Thus, only the phase shifts due to the path length difference between the illuminating beams remain.
The cancellation is not perfect due to errors in the reconstructed fields, which results in residual speckle noise in the interference pattern.

A salient feature of DHI is that the phase difference $\Delta \phi = \phi_2-\phi_1$ can be related to the three-dimensional shape of the scattering object~\cite{hildebrand1967multiple,abramson1976sandwich,pedrini1999shape, yamaguchi2001surface}.
Scattering from a flat surface, we would expect this phase difference to be $2\pi x/w_0$, where $x$ is the transverse spatial coordinate on the camera, and $w_0$ is the fringe period given by
\begin{equation}    
w_0 = \frac{\lambda}{2 \sin(\Delta \theta /2)} \approx \lambda / \Delta \theta,
\label{eqn:fringe_period}
\end{equation}
for small $\Delta \theta$.
However, if the illuminating beam propagates an additional distance $\delta z$ along the $z$ axis due to the presence of an object, it causes the countour fringes to shift by $\delta w = \delta z \tan(\theta) \approx  \delta z \theta$, leading to a phase shift of $\delta \phi = 2 \pi \delta w / w_0=\Delta \phi - 2\pi x/w_0$.
This shift is evident near the bottom of Fig.~\ref{fig:scattering_results}(d), where there is a discontinuity in the fringes due to the gap between the dice and the edge of the platform on which the dice is mounted.
There is also a slight chirp in the fringes along the dice faces due to their tilt.
We can map the phase difference pattern $\Delta \phi$ to the object depth using
\begin{equation}
    \delta z = \frac{\delta w}{\theta} = \frac{w_0 \delta \phi}{2 \pi \theta}  = \frac{\lambda}{2 \pi \theta \Delta \theta} \Delta \phi - \frac{x}{\theta}.
    \label{eqn:phase_to_depth}
\end{equation}
Using this equation, we obtain the depth map in Fig.~\ref{fig:scattering_results}(e), which is in agreement with the geometry of the scene [inset in Fig.~\ref{fig:scattering_results}(a)].
Each parameter in Eq.~\eqref{eqn:phase_to_depth} was obtained in the following way: (i) $\Delta \phi$ is obtained by computing the difference in the argument of the two reconstructed fields, as in Fig.~\ref{fig:scattering_results}(d), (ii) $\theta \sim 8 \times 10^{-2}$~rad is obtained by taking the ratio between $L_2$ (0.19 m) and $L_1$ (3.36 m), both measured with a tape measure, and finally, (iii) $\Delta \theta \sim 2 \times 10^{-4}$~rad is calculated using Eq.~\eqref{eqn:fringe_period} where $w_0$ is extracted from the period of the interference pattern in Fig.~\ref{fig:scattering_results}(d).
We also apply a median filter to the retrieved depth map in order to reduce residual speckle noise.

We briefly comment on the limitations of the depth resolution of the technique.
As indicated by Eq.~\eqref{eqn:phase_to_depth}, the sensitivity can be improved by increasing both $\theta$ and $\Delta \theta$.
However, an excessive illumination angle $\theta$ leads to shadows in the image where fringes cannot be resolved~\cite{hildebrand1967multiple}.
Moreover, $\Delta \theta$ cannot be increased arbitrarily since the camera pixel pitch $p$ sets the smallest resolvable fringe shift, i.e. $\Delta \theta_{\mathrm{max}} = 
\lambda/2p$, thus limiting the depth resolution to $\sim 2p/\theta$.
This technique can in principle be used to measure sub-mm depth features~\cite{pedrini1999shape,yamaguchi2001surface}.
\section{Conclusion and outlook}
To summarize, we showed that randomly phase-shifted interferograms can be coherently combined into a single intensity-correlation hologram.
This allowed us to extend the integration time of the hologram beyond the interferometer coherence time, thereby facilitating phase imaging in low-light and mechanically-unstable conditions.
Our phase-retrieval algorithm is fast, simple, and resilient to camera readout noise.
As a proof-of-principle demonstration of remote sensing, we determined the wavefront of a signal wave scattered from an object at a distance of $L_{\mathrm{obj}}\sim 3$~m.
We also used digital holographic interferometry to measure the 3D shape of an object with mm resolution.

In future work, one could extend $L_{\mathrm{obj}}$ to much longer distances (e.g. hundreds of meters).
Noise arising due to object movement and atmospheric turbulence can be overcome by capturing the intensity-correlation hologram on microsecond timescales using high-framerate or event-based cameras~\cite{thekkadath2023intensity}.
The object distance could even exceed the coherence length of the laser so long as the detector timing resolution is faster than the coherence time of the laser.
With an increasing distance but fixed object size, one would eventually need to mitigate diffraction limits by enlarging the telescope diameter $D$.
Moreover, the camera pixel pitch should be smaller than $\lambda f / D$ to resolve the interference pattern resulting from the highest frequency component of the imaging system.

\begin{backmatter}

\bmsection{Acknowledgments}
\noindent The authors thank Denis Guay and Doug Moffatt for their technical support. 

\bmsection{Disclosures}
\noindent The authors declare no conflicts of interest.

\bmsection{Data Availability}
\noindent The raw data may be obtained from the authors upon request.

\end{backmatter}

\bibliography{refs}

\end{document}